\documentclass[aoas,MSNbibl,nameyear,rotating,dvips]{arximspdf}
\usepackage{dcolumn}
\usepackage{graphicx}
\usepackage{url,breakurl}

\doi{10.1214/14-AOAS737} 
\volume{8}
\issue{3}
\pubyear{2014}
\firstpage{1538}
\lastpage{1560}
\docsubty{FLA}

\makeatletter
\newcolumntype{d}[1]{D{.}{.}{#1}}

\def\btheta{\bolds{\theta}}
\def\bphi{\bolds{\phi}}
\def\bpsi{\bolds{\psi}}
\makeatother

\begin{document}
\begin{frontmatter}

\title{Nonlinear predictive latent process models for integrating
spatio-temporal exposure data from~multiple sources}
\runtitle{Aggregating multiple data sets using a latent process}

\begin{aug}
\author[A]{\fnms{Nikolay}~\snm{Bliznyuk}\corref{}\ead[label=e1]{nbliznyuk@ufl.edu}\thanksref{T1,T2,M1}},
\author[B]{\fnms{Christopher J.}~\snm{Paciorek}\ead[label=e2]{paciorek@stat.berkeley.edu}\thanksref{T1,M2}},
\author[C]{\fnms{Joel}~\snm{Schwartz}\ead[label=e3]{jschwrtz@hsph.harvard.edu}\thanksref{T1,M3}}
\and
\author[D]{\fnms{Brent}~\snm{Coull}\ead[label=e4]{bcoull@hsph.harvard.edu}\thanksref{T1,M3}}
\runauthor{Bliznyuk, Paciorek, Schwartz and Coull}
\affiliation{University of Florida\thanksmark{M1}, University of
California\thanksmark{M2} and\\ Harvard School of Public
Health\thanksmark{M3}}
\address[A]{N. Bliznyuk\\
Department of Agricultural\\
\quad and Biological Engineering\\
University of Florida\\
406 McCarty Hall C\\
Gainesville, Florida 32611\\
USA\\
\printead{e1}} 
\address[B]{C.~J. Paciorek\\
Department of Statistics\\
University of California, Berkeley\\
367 Evans Hall\\
Berkeley, California 94720\\
USA\\
\printead{e2}}
\address[C]{J. Schwartz\\
Department of Environmental Health\\
Harvard School of Public Health\\
665 Huntington Avenue\\
Landmark Center Room 415\\
Boston, Massachusetts 02115\\
USA\\
\printead{e3}}
\address[D]{B. Coull\\
Department of Biostatistics\\
Harvard School of Public Health\hspace*{8pt}\\
655 Huntington Avenue\\
Boston, Massachusetts 02115\\
USA\\
\printead{e4}}
\end{aug}
\thankstext{T1}{Supported by Grants from the National Institute of Environmental Health Sciences
(ES07142, ES000002, ES012044, ES016454, ES009825), the National Cancer Institute (CA134294), and the U.S. Environmental
Protection Agency (EPA) Grant R-832416.}
\thankstext{T2}{Supported by the postdoctoral training Grant CA090301 from the National Cancer Institute.}

\received{\smonth{2} \syear{2012}}
\revised{\smonth{2} \syear{2014}}

%
\begin{abstract}
Spatio-temporal prediction of levels of an environmental exposure is an
important problem in environmental epidemiology. Our work is motivated
by multiple studies on the spatio-temporal distribution of mobile
source, or traffic related, particles in the greater Boston area. When
multiple sources of exposure information are available, a joint model
that pools information across sources maximizes data coverage over both
space and time, thereby reducing the prediction error.

We consider a Bayesian hierarchical framework in which a joint model
consists of a set of submodels, one for each data source, and a model
for the latent process that serves to relate the submodels to one
another. If a submodel depends on the latent process nonlinearly,
inference using standard MCMC techniques can be computationally
prohibitive. The implications are particularly severe when the data for
each submodel are aggregated at different temporal scales.

To make such problems tractable, we linearize the nonlinear components
with respect to the latent process and induce sparsity in the
covariance matrix of the latent process using compactly supported
covariance functions. We propose an efficient MCMC scheme that takes
advantage of these approximations. We use our model to address a
temporal change of support problem whereby interest focuses on pooling
daily and multiday black carbon readings in order to maximize the
spatial coverage of the study region.
\end{abstract}

%
\end{frontmatter}

\section{Introduction and background}\label{secintro}

An important scientific goal in environmental health research is the
identification of sources of air pollution responsible for
the well-documented health effects of air pollution. A pollution source
of great interest is motor vehicle (i.e., traffic) emissions.
Because traffic pollution is inherently higher near busy roads and
major highways and falls off to background levels relatively
quickly in space, concentrations of traffic-related pollutants exhibit
large amounts of spatial heterogeneity within an urban
area. Therefore, epidemiologic studies of the health effects of
traffic pollution that use a centrally sited ambient monitor suffer
from large amounts of exposure measurement error [\citet{zeger2000}]. However, because it is not always feasible to obtain
exposure recordings at each study subject's residence at a given time
(a special case of spatio-temporal misalignment), it is now common
practice in air pollution epidemiology for researchers to collect data
from monitoring networks on the intra-urban spatio-temporal
variability in traffic pollution levels. These data are used to make predictions
of the exposure process, which are then used as a surrogate for true
exposures in health effects models [\citet{adar2010}; \citet{berhane2004}; \citet{wanne2009}]. Note that this creates a
measurement error problem [\citet{gryp09}].

In this article we consider statistical models for prediction of
spatio-temporal concentrations of black carbon (BC), thought to be a
surrogate for traffic-related air particle levels [\citet{Janssen11}], in the greater Boston-area.
One complicating factor in the development of such models in our
Boston-area analysis, however,
is that the logistical and financial demands of maintaining a dedicated
monitoring network are prohibitive. Accordingly, rather than set up a
single network with
a standardized monitoring protocol, our collaborators have augmented existing
ambient monitoring data with targeted residence-specific
indoor pollution monitoring aimed at increasing both the spatial and temporal
coverage of the study region and period, respectively. Early work by
our group [\citet{gryp07}]
focused on latent variable models for the integration of
spatio-temporal data from multiple sources
when all data were measured at the same temporal (in this case, daily)
scale. The resulting number
of monitors producing daily BC data was modest (under 90), limiting our
ability to fully explore the spatio-temporal
structure in the data. Specifically, such data sparsity motivated us to
fit relatively simple
spatial models separately for warm and cold seasons, as opposed to
fitting more complex and likely
more realistic spatio-temporal correlation structures across the entire
study period.

Since this initial work, data at additional spatial locations have been
collected. In this work, we consider
data from 93 additional indoor monitors and explore how incorporation
of these data improves our ability
to explore the spatio-temporal patterns in the resulting monitoring
data and ultimately the predictive performance
of the resulting exposure models. One factor complicating the
integration of these new data is the fact that
this more recent monitoring campaign yielded concentration data at
temporal scales different than the original BC
data. Whereas the original data were collected on a daily time scale,
the more recent monitoring campaign yielded
multiday integrated readings. Therefore, our scientific interest
focuses on the integration of data from these disparate sources
into a unified exposure prediction framework, while rigorously
accounting for changes in temporal support and the fact
that different monitors operate irregularly in time. Given a modeling
strategy that satisfies these goals, we assess the improvement in
predictive performance of the models that incorporate all the data
versus simpler models that only use the original
daily data. While there has been a wide body of statistical work on
spatio-temporal modeling of air pollution, most of these efforts have
focused on data without substantial temporal misalignment and with a
single type of pollution measurement. Although there is a considerable
literature on the change of spatial and spatio-temporal support
[\citet{gelfand01}] and the use of aggregated data in spatial
statistics [\citeauthor{go02} (\citeyear{go02,go07}); \citet{fue05}],
these proposed methods largely rely on the linear change-of-support and
data assimilation. For example, \citeauthor{calder2007} (\citeyear{calder2007,calder2008}) develops \mbox{dynamic}
process convolution models---effectively, multivariate time series
models---for multivariate spatio-temporal air quality data that allow
one to solve the linear change-of-support problem. We are not aware of
references that focus on the nonlinear change of temporal support in
spatio-temporal statistics.

%
\begin{figure}

\includegraphics{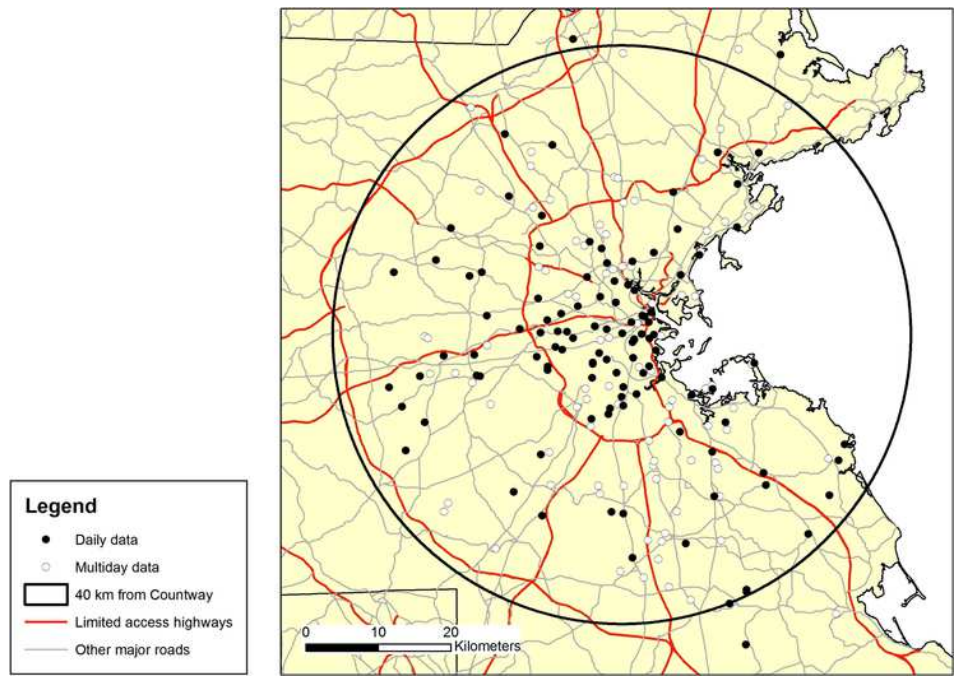}

\caption{Spatial coverage of monitors. The HSPH (Countway) monitor is
at the center of the circle in Downtown Boston.}\label{figmonitorsspat}
\end{figure}

We now outline the structure of the available BC data in more detail.
The three data types that we use in our
model and describe below are (i) \textit{daily} average outdoor BC
concentrations (abbreviated as BCO), (ii) \textit{daily} average indoor
BC concentrations (BCI) and (iii) \textit{multiday} aggregated indoor
BC concentrations (BCA), in $\mu g / m^3$. Figure~\ref{figmonitorsspat}~and Figure~4 of the online
supplements [\citet{suppl14}] display the spatial and temporal coverage of the study
region and periods, distinguishing the different data types.

\subsection*{Daily outdoor data (BCO)}
A sizeable fraction of the BCO readings that we use come from \citet{gryp07}.
These data, generated from two different exposure assessment studies,
were collected by outdoor monitors at 48 spatial locations in Boston
and its suburbs over the period from mid October 1999 to the end of
September 2004. The length of time each monitor operated ranged from 2
weeks to hundreds of days. These monitoring efforts resulted in 4219
daily BC readings. Our analyses supplement these data with additional
daily BC data collected as part of a recent NIH Program Project Grant
(PPG), which added 2696 daily readings from 52 distinct sites
taken between mid March 2006, and early November 2008. The observations
from the two studies do not occur at the same
spatial locations, thus, we have 100 distinct sites, with over 6900
daily BCO readings.

\subsection*{Daily indoor data (BCI)} These data consist of 318 daily indoor
concentrations of BC from 45 distinct households, recorded between mid
November 1999 and early December 2003. Of these 45 sites, 30
overlap spatially and temporally with the BCO data. Further
details are in \citet{gryp07}.

\subsection*{Multiday aggregated indoor data (BCA)} Multiday measurements of
indoor BC were collected as part of the Normative Aging Study (NAS).
There are 93 observations, one
per household, each of which is a measurement of concentration
aggregated over multiple days; the corresponding daily concentrations
are not available.
The lengths of measurement, which range from 3 to 14 days, and starting
dates of the monitoring periods are different across the
households. The data correspond to the period from mid July 2006
through late March 2008. The spatial locations of the multiday data are
distinct from those of the daily data.

To achieve the scientific goals outlined above, we develop a
Bayesian hierarchical framework for inference and prediction where a
joint model for all exposure measurements depends on a set of
submodels, one for each data source, and a model for the latent
process that relates the submodels to one another. In particular, we
focus on the case in which the submodels depend on the latent process
nonlinearly, which frequently occurs when the different data sources
yield data on different surrogates of pollution or at varying temporal
or spatial scales.

Inference for nonlinear hierarchical models with latent Gaussian
structure is
computationally challenging. When the likelihood is not Gaussian,
likelihood-based and Bayesian inference involve high-dimensional
integration with respect to the random effects that cannot be
expressed in closed form. While MCMC is a standard approach for such
models in a Bayesian framework, convergence and mixing are often
troublesome [\citet{chriswaag2002}, \citet{christrob2006}]
because of the high-dimensionality of the random effects
and the dependence between random effects (particularly in
spatio-temporal specifications) and cross-level dependence between
random effects and their hyperparameters [\citet{rueheld}, \citet{rueinla}].

Our main methodological contribution is development of an efficient,
yet straightforward, MCMC algorithm for Bayesian inference on model
parameters and prediction of arbitrary functions of the latent
process. Within our hierarchical model, it is based on the
approximation of
nonlinear regression functions by ``linearizing'' them with respect to
the latent process values over the region of their high posterior
probability.

The paper is organized as follows: in Section~\ref{secstat}
we describe the overall hierarchical modeling strategy, proposing a
nonlinear statistical model in
Section~\ref{subsecnonlin} that is approximated through
``linearization.'' In
Section~\ref{seccasecomputational} we present a computational
strategy to reduce the cost of Bayesian inference and prediction and
discuss the relative merits of our approach and existing approximation
methods. Section~\ref{secresults} is devoted to model selection and
validation for pollutants in the greater Boston data, assessment of the
adequacy of
linearization and the results of Bayesian inference and
prediction. Discussion and concluding remarks are in
Section~\ref{secdiscussion}. Technical details and supplementary
figures and tables are in the online supplements [\citet{suppl14}].

\section{Statistical model}\label{secstat}
This section defines the joint model for the observed data.
The individual models for the observations of each type are linked
through the latent process.
The nonlinear model for the multiday data is subsequently
``linearized'' for the sake of computational
tractability.

\subsection{Nonlinear observation model}\label{subsecnonlin}

Following \citet{gryp07}, the latent spatio-temporal process,
$\eta$,
is a proxy for the logarithm of the true daily average concentration of outdoor
black carbon (BC). The model for $\eta$ will be specified in the next
subsection.
For notational simplicity,
we will often abbreviate the space--time indices using subscripts,
for example, $\eta(s_i,t_j)$ as $\eta_{ij}$ for the value of the latent
process at site $s_i$ on date $t_j$.
The logarithms of the observed outdoor and indoor daily average BC
concentrations, $Y_{ij}^O$ and $Y_{ij}^I$,
are related to the latent process as
\begin{eqnarray}
\label{eqYO} Y_{ij}^O &=& \eta_{ij} +
\varepsilon_{ij}^O,
\\
\label{eqYI} Y_{ij}^I &=& \alpha_{0i} +
\alpha_{1I} \eta_{ij} + \varepsilon_{ij}^I,
\end{eqnarray}
where $\{\alpha_{0i}\}$ are household-specific fixed effects
and $\varepsilon_{ij}^O, \varepsilon_{ij}^I$ are instrument errors.
The household-specific effects are introduced as in \citet{gryp07}
in order to account for that differences in
penetration efficiencies of particles that depend
on properties of the building. In the absence of instrument error,
setting the slope $\alpha_{1I} = 1$ corresponds to the indoor BC being
proportional to the outdoor BC on the original scale, with the
proportionality constant $\exp(\alpha_{0i})$. The values of the slope
$\alpha_{1I}$ less than one---such as those observed with our data---allow one to account for the slower than linear increase in the indoor
BC as the outdoor BC grows, relative to the proportional concentration model.

The model for the observed average \textit{multiday}
concentration of indoor black carbon at a site $s_i$ is defined as
\begin{equation}
\label{eqYSnonlin} Y_i^A = \alpha_{0i} +
g_i\bigl(\vec{\eta}_i^A\bigr) +
\varepsilon_i^A, 
\end{equation}
where $\vec{\eta}_i^A$ is the vector of (daily) latent process
values upon which the aggregate average reading at $s_i$ depends
and $\varepsilon_i^A$ is the instrument error.
We assume that the instrument error
processes
$\{\varepsilon_{ij}^O\}$, $\{\varepsilon_{ij}^I\}$ and $\{\varepsilon_i^A\}$ are
mutually independent Gaussian white noise with zero mean and variances
$\sigma_O^2$, $\sigma_I^2$ and $\sigma_A^2$, respectively.

Without loss of detail, let $Y_i^A$ be the logarithm of the sum (as
opposed to an average) of consecutive daily average concentrations of
indoor black carbon at site~$s_i$. The nonlinear regression model for
the multiday data is
\begin{equation}
\label{eqgi} g_i\bigl(\vec{\eta}_i^A
\bigr) = \log\sum_j \exp(\alpha_{1I}
\cdot\eta_{ij}).
\end{equation}
The nonlinearity arises because the multiday readings are aggregated
on the original rather than on the logarithmic scale. For instance,
without the instrument error, that is,
if $\varepsilon_{ij}^I = \varepsilon_i^A = 0$, $Y_i^A$ would be
the logarithm of $\sum_j \exp(Y_{ij}^I)$, the sum of consecutive
daily readings of (daily) average indoor black carbon
concentrations at site $s_i$. Trivially, equation~(\ref{eqYI})
is a special case of equation~(\ref{eqYSnonlin}).

Note that there is only a single reading $Y_i^A$ for each
household, so the home-specific intercepts $\alpha_{0i}$ are not
identifiable in the model of equation~(\ref{eqYSnonlin}). We therefore
absorbed them into $\varepsilon_i^A$, but introduced the parameter $\alpha_{00}$
to capture the \mbox{population} intercept. Exploratory
analysis revealed that the slope parameter $\alpha_I$
can be significantly different for the models
for $Y^I$ and $Y^A$.
Consequently, the model~(\ref{eqYSnonlin}) was changed to
\begin{equation}
\label{eqYSnonlinid} Y_i^A = \alpha_{00} + \log\sum
_j \exp(\alpha_{1A} \cdot
\eta_{ij}) + \varepsilon_i^A.
\end{equation}
The coefficient $\alpha_{1A}$ is allowed to be different from
$\alpha_{1I}$ in the model for $Y^I$, in order to account for
(i) data aggregation and rounding errors, since the monitors from
the NAS study do not run for an integer number of days, and
(ii) demographic differences in households since the multiday
data come from a study (targeting elderly people) different from
the study providing the daily indoor data.

\subsection{Latent process model}
\label{seccasemodelmain}
\label{seccasemodel}
\label{subsubidlatent}

The latent process
at site $s_i$ on day $t_j$ is modeled~as
%
\begin{equation}
\label{eqeta} \eta(s_i,t_j) = x(s_i,t_j)^{\mathsf{T}}
w_x + \zeta(s_i,t_j) +
u(s_i,t_j), 
\end{equation}
where $x(s_i,t_j)$ is a vector of observable predictors and
$\zeta(s_i,t_j) + u(s_i,t_j)$ accounts for unobservable spatio-temporal
variability.
In order to ensure identifiability, we let $\zeta$ capture the
temporally long-range spatio-temporal variability and
$u$ capture the temporally short-range variability.
Equivalently, for a fixed value $s_0$ of~$s$, $u(s_0,\cdot)$ is
a stationary temporal process with rapidly decaying dependence, and
$\zeta(s_0, \cdot)$ is a long-range temporal process, possibly with
nondecaying dependence.

%
\begin{sidewaystable}
\tabcolsep=0pt
\tablewidth=\textwidth
\caption{Posterior summaries of the coefficients of the
observed predictors under model $M(U=1,\mathit{GST}=0,A=1)$}\label{tab4}
\begin{tabular*}{\tablewidth}{@{\extracolsep{\fill}}@{}lcd{2.3}d{2.3}d{2.3}d{2.3}@{}}
\hline
$\bolds{w_i}$ & \textbf{Predictor} & \multicolumn{1}{c}{\textbf{Mean}}
& \multicolumn{1}{c}{\textbf{2.5\%}} & \multicolumn{1}{c}{\textbf{50\%}} & \multicolumn{1}{c@{}}{\textbf{97.5\%}}\\
\hline
$w_{1}$ & \texttt{1}, the intercept & 4.580 & -2.190 & 4.176 & 13.516 \\
$w_{2}$ & \texttt{log\_pop\_sqkm}, log of population per square km &0.259 & 0.021 & 0.262 & 0.494 \\
$w_{3}$ & \texttt{log\_adtxlth100m}, log of traffic density & -0.177 &-0.306 & -0.176 & -0.049 \\
$w_{4}$ & \texttt{nlcd}, land use index & \multicolumn{1}{c}{$2.65\cdot10^{-4}$} & \multicolumn{1}{c}{$1.19\cdot10^{-4}$} & \multicolumn{1}{c}{$2.63\cdot10^{-4}$} & \multicolumn{1}{c@{}}{$4.14\cdot10^{-4}$}\\
$w_{5}$ & \texttt{loghsph}, log of HSPH monitor readings & 0.767 & 0.742& 0.767 & 0.793 \\
$w_{6}$ & \texttt{wind\_sp}, wind speed & 0.129 & 0.014 & 0.130 & 0.244 \\
$w_{7}$ & \texttt{log\_pbl}, log of planetary boundary layer & -0.073 &-0.244 & -0.074 & 0.095 \\
$w_{8}$ & \texttt{log\_pop\_sqkm * wind\_sp} & -0.028 & -0.053 & -0.028 &-0.003 \\
$w_{9}$ & \texttt{log\_adtxlth100m * wind\_sp} & 0.002 & -0.011 & 0.002 &0.015 \\
$w_{10}$ & \texttt{log\_pbl * wind\_sp} & -0.023 & -0.040 & -0.023 &-0.006 \\
$w_{13}$ & \texttt{log\_pop\_sqkm * log\_pbl * wind\_sp} & 0.004 & 0.001& 0.005 & 0.008 \\
$w_{14}$ & \texttt{log\_adtxlth100m * log\_pbl * wind\_sp} & 0.000 &-0.002 & 0.000 & 0.002 \\
\hline
\end{tabular*}
\end{sidewaystable}

For our case study, components of the vector of observable covariates
$x(s,t)$ in equation~(\ref{eqeta}) are provided in Table~\ref{tab4}.
They include (i) spatially-varying variables---population density,
traffic density and land use; (ii)\break temporally-varying variables---readings from the Harvard School of Public Health (HSPH) central site monitor,
meteorological variables (wind speed and planetary boundary layer);
and (iii) interaction terms. We use the logarithm of readings from
the HSPH central site monitor as a predictor rather than as a response
in order to enable comparisons with earlier work of \citet{gryp07} that set $u=0$. The implication is that much of the temporal
variability common to all sites is captured by observations from the
central site and that the temporal components of the model capture
variability above and beyond that measured at the central site.

Following \citet{owy}, we let $\zeta$ capture the
long-range spatio-temporal variation, often referred to as the unknown
smooth spatio-temporal trend. In the spirit of \citet{wang98}, we use
penalized splines, so that the trend can be represented as
\begin{equation}
\zeta(s,t) = z(s,t)^{\mathsf{T}} w_z,
\end{equation}
where $z(s,t)$ is a column vector of known basis functions evaluated at
$(s,t)$ and $w_z$ is a column vector of the corresponding coefficients.
We define the actual form of $z(s,t)$ and constraints on $w_z$ below.
Because the spatio-temporal coverage by the monitors is sparse (about
7300 observations from over 2700 distinct days and at most 200
sites), unconstrained spatio-temporal smoothing would be unreliable in
parts of the domain without observations. Instead we put constraints on
the spatio-temporal smoother by requiring the smoother to be periodic,
thereby borrowing strength across years when estimating the trend. This
also allows one to make predictions outside the temporal range of the
observations. The local deviations of the latent process from the
periodic term will be accounted for by the $u(s,t)$
process.

We decompose the long-range spatio-temporal trend as
\[
\zeta(s,t) = g_S(s) + g_T(t) + g_{\mathit{ST}}(s,t),
\]
where $g_S$ and $g_T$ are smooth functions of spatial coordinates and
of time,
respectively, and $g_{\mathit{ST}}$ is a function representing the long-range
(in time) spatio-temporal interaction.
Here, $g_T$ is the annual (cyclic) temporal trend,
so that $g_T(t) = g_T(d_t)$, where $d_t = \operatorname{mod}(t,365)$ is the day of the
year if leap years are ignored. We use a thin-plate spline with 60
knots to model $g_S$, a cubic spline with seven equally spaced knots to
model $g_T$, and
the tensor product of spatial and temporal basis functions to model the
interaction, $g_{\mathit{ST}}$ [\citet{wood}]. To ensure that the temporal trend
is periodic, continuous and differentiable at $t=0$, linear constraints
were placed on the coefficients of $g_T$ and of $g_{\mathit{ST}}$; see online
supplements [\citet{suppl14}], Section~A.5.3.
Thus, the model of equation~(\ref{eqeta}) can be written as a linear model
\begin{equation}
\label{eqbco2} \eta(s,t) = c(s,t)^{\mathsf{T}} w + u(s,t),
\end{equation}
where
\[
c(s,t)^{\mathsf{T}} = \bigl\{x(s,t)^{\mathsf{T}},\bigl[s;\bphi(s)
\bigr]^{\mathsf{T}}, \bigl[t;\bpsi(t)\bigr]^{\mathsf{T}}, \bigl[s;\bphi(s)
\bigr]^{\mathsf{T}} \otimes\bigl[t;\bpsi (t)\bigr]^{\mathsf{T}}\bigr\}
\]
is a row vector of ``predictors'' and
\[
w = \bigl[w_x^{\mathsf{T}}, w_S^{\mathsf{T}},
w_T^{\mathsf{T}}, w_{\mathit{ST}}^{\mathsf{T}}
\bigr]^{\mathsf{T}}
\]
is a column vector of coefficients. Here, the $i$th component of $\bphi
(\cdot)$ is
$\phi_i(\cdot) = \phi(\cdot,s^{(i)})$, the spatial basis function
centered at the knot $s^{(i)}$; similarly, $\psi_j(\cdot) = \psi(\cdot,
t^{(j)})$ is the $j$th temporal\vspace*{1pt} basis function centered at the knot
$t^{(j)}$, for example, $\psi_j(d_t)=|d_t-t^{(j)}|^3$.
Following \citet{wood}, we penalize the
square of the second derivative of the nonparametric smooth terms to
prevent overfitting. This approach is attractive because the penalty
matrices for $g_S$
and $g_T$ can be written as symmetric positive semidefinite quadratic forms
in $w_S$ and $w_T$. For example, the penalty for $g_T$ is
\begin{equation}
\label{eqpenaltyt} \mathcal{P}_T = \int\bigl\{g_T''(t)
\bigr\}^2 \,dt = w_T^{\mathsf{T}} \cdot M_T
\cdot w_T
\end{equation}
for some symmetric positive semidefinite matrix $M_T$. The spatial and
temporal marginal penalty matrices $M_{\mathit{ST},S}$ and $M_{\mathit{ST},T}$ for the
smooth interaction term $g_{\mathit{ST}}$ are derived in the online supplements [\citet{suppl14}],
Section~A.5.2.
These penalty matrices are subsequently used to define a precision
matrix for the multivariate normal prior on $w$ as the Bayesian
analogue of
the penalized log-likelihood criterion with
penalty matrices $M_S$, $M_T$, $M_{\mathit{ST},S}$ and $M_{\mathit{ST},T}$ [\citet{rwc}]. This prior has a zero mean and precision matrix
\begin{equation}
\label{eqpriorw} Q_w = \operatorname{blkdiag} \biggl\{\Delta\cdot I_{\dim(w_x)},
\frac{M_S}{\tau_S^2}, \frac{M_T}{\tau_T^2}, \frac{M_{\mathit{ST},S}}{\tau_{\mathit{ST},S}^2} + \frac
{M_{\mathit{ST},T}}{\tau_{\mathit{ST},T}^2} \biggr
\},
\end{equation}
where a small multiple $\Delta$ of the identity matrix is used to
ensure that the prior on the linear coefficients $w_x$ is proper and
where $\operatorname{blkdiag}$ is a block-diagonal matrix
with blocks listed as arguments.

We use a Gaussian process model for $u$ in order to account for the
short-range temporal variability and spatio-temporal interaction.
Given the data sparsity, we model the covariance function for $u$ in a
separable fashion for simplicity as
\begin{equation}
\label{eqepscovf} \operatorname{Cov}\bigl\{u(s,t), u\bigl(s',t'
\bigr)\bigr\} = 
\sigma_u^2 \cdot
C_S\bigl(s,s'|\theta_S\bigr) \cdot
C_T\bigl(t,t'|\theta_T\bigr),
\end{equation}
where $C_S$ and $C_T$ are spatial and temporal correlation
functions.

To model spatial dependence, we use the Mat\'ern family of correlation
functions
\begin{equation}
\label{eqmatern} C_S(s,s+h) = \bigl(2 \sqrt{\nu} \theta_S \|h
\|_2\bigr)^\nu\cdot K_\nu\bigl(2\sqrt{\nu}
\theta_S \|h\|_2 \bigr)/\bigl\{2^{\nu-1} \Gamma(\nu)
\bigr\},
\end{equation}
where $\nu, \theta_S > 0$, $\Gamma(\cdot)$ is the gamma function and
$K_\nu(\cdot)$ is the modified Bessel function of order $\nu$
[\citet{baner}].
The smoothness parameter $\nu$ is difficult to estimate accurately
unless the
spatial resolution of the data is very fine [\citet{gn12}].
Due to the spatial sparsity of the set of monitors, we hold $\nu$ fixed
at 2,
thereby representing smooth short-range (based on the tapering
described next) variation, in $u(s,t)$.
Nonsmooth variability is accounted for by the errors $\varepsilon$.

{Examination of the plots} of autocorrelation and partial
autocorrelation functions for residuals from a monitoring station with a
long series of daily measurements suggested that temporal dependence
can be explained well by an order-one autoregressive process with
moderate lag-one correlation (of less than 0.5).
Under the plausible assumption that the rates of decay of the
temporal autocorrelation are similar across all monitoring stations, it can
be seen that the components of $u$ that are 7 days or more apart are
practically uncorrelated since the correlation is less than $10^{-2}$.
Consequently,
we introduce sparsity structure into the covariance matrix explicitly
via {covariance tapering} [\citet{fu06}]. As a temporal
correlation function, we use
the product of the exponential and the (compactly supported) spherical
correlation functions
\begin{equation}
\label{eqtempcorr} C_T(t, t+h) = \exp(-\theta_T \cdot h)
\cdot \max\bigl\{(1-h/r),0\bigr\}^2 \bigl\{1 + h/(2r)\bigr\}
\end{equation}
for $r=7$, which behaves similarly to the exponential correlation
function when $h$ is small, and is exactly zero when $h \geq7$.
The benefits of tapering for the computational aspects of Bayesian
inference will be discussed Section~\ref{seccasetaper}.

\subsection{Linearized observation model}\label{subseclin}

We will refer to the set of
equations \mbox{(\ref{eqYO})--(\ref{eqeta})}
as
the \emph{nonlinear model}. MCMC for such models can be very
inefficient, if tractable at all. For example, if one puts a Gaussian
spatio-temporal process prior on $u$, one needs to sample from a
nonstandard density for the vector of latent process values (here, of
dimension 712) that enters the nonlinear model for $Y^A$. The values
cannot be analytically integrated over in the joint model.
In this subsection
we develop the idea of ``linearization'' of the nonlinear regression
function of
equation~(\ref{eqgi})
about some ``central'' value $\eta^{A*}$ of the latent process and
briefly discuss the practical choices for $\eta^{A*}$.

\subsubsection{Linearization}\label{subsublin}

The linearized model is obtained by doing a Taylor series expansion
of the nonlinear regression function $g_i$ in equation~(\ref{eqgi})
about some
``central'' value $\vec{\eta}_i^{A*}$ of vector $\vec{\eta}_i^A$:
\begin{equation}
\label{eqYSlin} Y_i^A = G_i(
\alpha_{1A}) + \alpha_{00} + \sum
_{j=1}^{J_i} b_{ij}(\alpha _{1A})
\eta_{ij} + \varepsilon_i^A,
\end{equation}
where $J_i$ is the number of days in the aggregated reading at site
$s_i$, $J_i \in\{3,\ldots,14\}$.
Here, $b_{ij}$ and $G_i$ are known deterministic functions of $\alpha_{1A}$:
\begin{eqnarray}
G_i(\alpha_{1A}) &=& g_i\bigl(\vec{
\eta}_i^{A*}\bigr) - 
\bigl\{b_i(
\alpha_{1A})\bigr\}^{\mathsf{T}} \vec{\eta}_i^{A*},
\\
b_i(\alpha_{1A}) &=&
[b_{i1}, \ldots, b_{iJ_i}]^{\mathsf{T}} = \frac{\partial g_i(x)}{\partial x}
\bigg| _{x=\vec{\eta
}_i^{A*}}\quad\mbox{and}
\\
\vec{\eta}_i^{A} &=& [\eta_{i1}, \ldots,
\eta_{iJ_i}]^{\mathsf{T}},
\\
\frac{\partial g_i(\vec{\eta}_i^A)}{\partial\eta_{ij}} &=& \frac{\alpha_{1A} \exp(\alpha_{1A} \eta_{ij})}{ \sum_j \exp(\alpha_{1A}
\eta_{ij})}. %
\end{eqnarray}
Notice that the model obtained by replacement of equation~(\ref
{eqYSnonlin}) by equation~(\ref{eqYSlin}) is a {conditionally
linear model} given $\alpha_{1A}$.

Define $v = (w; \{\alpha_{0i}\})$ and let $\gamma$ be
the vector of all remaining parameters, which includes
$\alpha_{1I}, \alpha_{1A}, \sigma_O^2, \sigma_I^2, \sigma_A^2$,
variance components controlling the smoothness of~$\zeta$
and parameters of the covariance function of $u$.
We can then write the linearized joint model for the observed data of
all types in matrix form as
\begin{equation}
\label{eqYjoint} Y = H(\alpha_1) \cdot(1; v) + \xi,
\end{equation}
where $\xi= X(\alpha_1)u + \varepsilon$ and $\alpha_1 = (\alpha
_{1I},\alpha_{1A})$. Here, $H$ and $X$ are matrices that do not depend
on $v$, as follows from equations~(\ref{eqYI})--(\ref{eqeta}) and (\ref{eqYSlin}).
The scalar $1$ is necessary to capture the
offset due to $G_i(\alpha_{1A})$ in the linearized model for
$Y^A$, equation~(\ref{eqYSlin}). Notice that, conditional on $\alpha
_1$, this is a linear model with dependent Gaussian errors, which
allows a computationally efficient implementation of an MCMC sampler,
discussed in Section~\ref{seccasecomputational}.

\subsubsection{Choice of the central value of the latent process}

The scheme outlined above assumes the availability of
the point $\eta^{A*}$ about which the linearization
is performed. In this section we detail how this value
can be obtained and justified. We use a standard bracket notation for
marginal, $[\cdot]$, and conditional, $[\cdot|\cdot]$, densities
[\citet{rwc}].

Upon defining $Y^{\mathit{OI}} = (Y^O,Y^I)$ and changing the order of
conditioning as
\begin{eqnarray*}
&& \bigl[Y^{\mathit{OI}},Y^A|\eta^A,w,\gamma\bigr] \bigl[
\eta^A,w,\gamma\bigr] 
\\
&&\qquad = \bigl[Y^A|
\eta^A,w,\gamma,Y^{\mathit{OI}}\bigr] \bigl[\eta^A|w,
\gamma,Y^{\mathit{OI}}\bigr] \bigl[w,\gamma |Y^{\mathit{OI}}\bigr]
\bigl[Y^{\mathit{OI}}\bigr],
\end{eqnarray*}
%
it is seen that the posterior $[\eta^A, w,\gamma|Y^{\mathit{OI}}]$ for the daily
data, $Y^O$ and $Y^I$, implicitly acts as an informative prior
for the parameters and the latent process in the multiday model likelihood
$[Y^A|\eta^A,w,\gamma, Y^{\mathit{OI}}]$. (Since $\{\alpha_{0i}\}$ can be
integrated out analytically, $v$ is replaced by $w$ here.) Because
$[\eta^A, w,\gamma|Y^{\mathit{OI}}] = [\eta^A|w,\gamma, Y^{\mathit{OI}}][w,\gamma|Y^{\mathit{OI}}]$,
the mass of the density of the latent process vector $\eta^A$ is concentrated
around the best linear unbiased predictor (BLUP) $E(\eta^A|Y^{\mathit{OI}}, w =
\widehat{w}, \gamma= \widehat{\gamma})$,
where $\widehat{w}$ and $\widehat{\gamma}$ are some ``central''
values of $w$ and $\gamma$. This suggests the use of
$\eta^{A*} = E(\eta^A|Y^{\mathit{OI}}, w = \widehat{w}, \gamma= \widehat{\gamma})$
in the linearization. In fact, as the daily data become dense in space,
infill asymptotics suggest that the BLUP $E(\eta^A|Y^{\mathit{OI}}, w = \widehat
{w}, \gamma= \widehat{\gamma})$
converges to the true unobserved value of $\eta^A$. Consequently, (\ref
{eqYSlin})
provides a likelihood for $Y^A$ that results in Bayesian inferences and
predictions
that are asymptotically equivalent to those under the true nonlinear model.
Of course, the validity of this large-sample argument may be
questionable in
some applications. For our case study, we justify use of the linearized
model for BCA empirically using a cross-validation study in Section~\ref{subpred}. In Section~\ref{sublinvsnonlin} we assess the accuracy
of inferences under the linearized model against those under the
nonlinear model in the simplest case when neither long- nor short-range
dependence is included in the model.

Taylor expansion about
$\eta^{A*} = E(\eta^A|Y^{\mathit{OI}}, w = \widehat{w}, \gamma= \widehat{\gamma})$
is computationally tractable because the marginal posterior $[\gamma
|Y^{\mathit{OI}}]$ or the profile posterior $\sup_w[\gamma,w|Y^{\mathit{OI}}]$ can be
obtained analytically (up to a constant of proportionality) and hence
maximized efficiently to get $\widehat{\gamma}$; the corresponding
value of $w$ is available analytically. In\vspace*{1pt} contrast, a possible
alternative of expanding about the mode of $[\eta^A,w,\gamma|Y^{\mathit{OIA}}]$
would require a costly numerical optimization run.

Notice that the na\"{\i}ve solution to the temporal change of support
problem, that~is,
\begin{equation}
\label{eqref2} Y_i^A = \beta_0 +
\beta_1 \sum_{j=1}^{J_i}
\eta_{ij} + \varepsilon_i^A,
\end{equation}
arises as a special case of our linearized model when $\vec{\eta
}_i^{A*}$ is set to zero.
In this case, $G_i(\alpha_{1A}) = \log J_i$ and $b_{ij}(\alpha_{1A}) =
\alpha_{1A}/J_i$, where $J_i$ is the observation period length at the
site $s_i$. The linearized model becomes
\[
Y_i^A = \log J_i + \alpha_{00} +
\frac{\alpha_{1A}}{J_i}\sum_{j=1}^{J_i}
\eta_{ij} + \varepsilon_i^A,
\]
which is equivalent to the above ``naive'' model when the observation
period lengths $J_i$ are all equal. However, this is hardly appropriate
in our case study since the observation periods lengths vary from 3 to
14 days, which implies that $\alpha_{00}$ and $\alpha_{1A}$ cannot be
identified from $\beta_0$ and $\beta_1$. More importantly, the na\"{\i
}ve linearization about 0 is inferior from the methodological
standpoint since, unlike the linearization about
$E(\eta^A|Y^{\mathit{OI}}, w = \widehat{w}, \gamma= \widehat{\gamma})$, the
approximation error in the Taylor expansion does not go to zero as the
spatial design becomes dense.


\section{Computational considerations for Bayesian inference by MCMC}\label{seccasecomputational}

In this section we develop three strategies that lower the
computational burden of model fitting and prediction: (i) covariance tapering,
(ii) strategies for sampling from the posterior density of the model
parameters, and (iii) sampling strategies for latent process prediction.

\label{seccasetaper}

Without tapering, the covariance matrix of the vector $\xi$ in
equation~(\ref{eqYjoint}), $\Sigma_Y$, is numerically dense. It can
take on the order of several seconds on a modern computer to form and
factorize this matrix, making a long MCMC sample computationally
expensive. Tapering reduces the proportion of non\-zero entries
(the fill) of $\Sigma_Y$ to less than 2\%. In addition, we reorder
the observed data $Y^O$ lexicographically with respect to the temporal
index, which makes unnecessary the formal element reordering
approaches [\citet{fu06}]. This makes $\Sigma_Y$ a
{banded (block) arrowhead matrix} (see Figure~5 in the
online supplements [\citet{suppl14}] for a visualization), which
yields a very efficient sparse Cholesky factorization [\citet{govl}]. As a result,
the cost to evaluate the likelihood drops by at least an order of
magnitude.
For a general nonlinear model in which the joint posterior density of
$(v,\gamma)$ is computationally expensive to evaluate and tapering is
not appealing, our linearization strategy can be supplemented by the
dimension reduction scheme of \citet{blizn11} for efficient
approximation of high-dimensional densities.

\label{seccasecomputationalmcmc}

We now discuss a strategy for sampling from the posterior density of
model parameters.
Recall from Section~\ref{subsublin} that $\alpha= \{\alpha_{0i}\}$,
$v = (w; \alpha_0)$
and $\gamma$ is the vector of all other parameters.
We analytically
integrate $v$ from the model as $[v|\gamma,Y]$ is multivariate normal.
Consequently, we draw from $[v,\gamma|Y]$ using \emph{composition
sampling}, that is,
by sampling $\gamma^{(i)}$ from $[\gamma|Y]$, and then
by exactly sampling $v$ from $[v|Y,\gamma= \gamma^{(i)}]$, which is in
the spirit of the partially collapsed Gibbs samplers work of \citet{vandyk08}.
In order to sample from
$[\gamma|Y]$, we use an adaptive random walk Metropolis--Hastings (RWMH)
sampling scheme, in the spirit of \citet{haario}, that calibrates the
covariance matrix of the proposal distribution based on the past
trajectory of the Markov chain.
The lag-1 autocorrelation in the
components of $\gamma$ in the actual sampling was below 0.95, while mixing
for the components of $v$ was considerably better; see
Section~\ref{sublinvsnonlin} for details.
The actual expressions for $[\gamma|Y]$ and $[v|\gamma,Y]$
are provided in the online supplements [\citet{suppl14}], Section~A.1.

\label{subseccomppredict}

For health effects studies and for fine visualization of the spatio-temporal
variability of the latent process, one often needs to predict the
values of
the latent process, $\eta^P$, at a large set of spatio-temporal
indices, say, at
a regular grid with $N_s$ spatial sites over the course of $N_t$ days.
In order to simulate from $[\eta^P|Y]$ under the linearized model, one
needs to
(i) sample from $[\gamma,v|Y]$ as in Section~\ref{seccasecomputationalmcmc} and (ii)
for each state in the $(\gamma,v)$-chain, sample exactly from
$[\eta^P|\gamma, v, Y]$, which is a multivariate normal density.
If $(\gamma^*, v^*)$ is a given value of $(\gamma,v)$ and
$\operatorname{Cov}(u^P,Y|v^*,\gamma^*)=\Sigma_{u^P,Y}(\gamma^*)$ and $\operatorname{Var}(Y|v^*,\gamma^*)=\Sigma_{Y,Y}(\gamma^*)$, one generally needs to
efficiently compute
\begin{eqnarray*}
&& E\bigl(\eta^P|Y,v=v^*,\gamma=\gamma^*\bigr)
\\
&&\qquad  = E\bigl(
\eta^P|v=v^*,\gamma=\gamma^*\bigr)
\\
&&\qquad\quad{} + \Sigma_{u^P,Y}\bigl(
\gamma^*\bigr)\cdot\Sigma_{Y,Y}^{-1}\bigl(\gamma^*\bigr)\cdot
\bigl\{ Y-E\bigl(Y|v=v^*,\gamma=\gamma^*\bigr)\bigr\}.
\end{eqnarray*}
For example, if one estimates $E(\eta^P|Y)$ by Monte Carlo via
``Rao--Blackwelliza\-tion'' [e.g., \citet{roca}], then\vspace*{2pt} $E(\eta
^P|Y) \approx M^{-1} \sum_{i=1}^M E(\eta^P|\break Y,v=v^{(i)},\gamma=\gamma
^{(i)})$. ``Poor man's'' approximations of the form $E(\eta
^P|Y,v=v^*,\gamma=\gamma^*)$, where
$(v^*,\gamma^*)$ is the posterior mode or the posterior mean, are also possible.
Section~A.2 of the online supplements [\citet{suppl14}] provides computational
details of evaluation of $E(\eta^P|Y,v,\gamma)$ and of sampling from
$[\eta^P|v,\gamma, Y]$.


\section{Analysis and results for the greater Boston data}\label{secresults}

\subsection{Candidate models}

Here we consider whether simpler models, such as the model of \citet{gryp07}, achieve comparable predictive accuracy to the full model
presented in Section~\ref{secstat}.
We examine eight candidate models, each determined by a combination
of following 3 indicator variables:
$U$---whether the model includes a Gaussian process model for
the short-range dependence term, $u$, or assumes that $u=0$;
$\mathit{GST}$---whether an extra term $g_{\mathit{ST}}$ for
the smooth long-range spatio-temporal interaction is included; and $A$---whether the aggregated multiday data, $Y^A$, are used (so as to
assess their importance in improving predictions). We use this labeling
scheme to abbreviate the models, for example, $M(U=0, \mathit{GST}=0,A=1)$. We
assess the models through cross-validation with spatially
nonoverlapping subsets.

\subsection{Assessment of predictive performance on validation data}\label{subpred}
We allocated a total of 1593 daily outdoor black carbon readings from
48 distinct sites into four disjoint groups of 12 sites, with each
group having roughly 400 data values.
To achieve this, we generated random partitions of the 48 sites into 4
groups many times and chose the partition that maximized the minimum
pairwise distance between sites and achieved roughly the same number of
observations in each group.
We held out each of the four validation subsets in turn, training the
model with the remaining observations and obtaining predictions to
compare with the held-out subset. Although the training and validation
subsets of data are spatially nonoverlapping, they are not temporally
disjoint. To expedite model fitting, we used optimization to find the
mode $\widehat{\gamma}$ of $[\gamma|\operatorname{train}]$ and then analytically
obtained the \mbox{corresponding} value $\widehat{v}(\widehat{\gamma})$ that
maximizes $[v|\gamma=\widehat{\gamma},\operatorname{train}]$,
after which we use the (empirical) BLUP
$E(Y^V|\operatorname{train}, v=\widehat{v},\gamma=\widehat{\gamma})$ to obtain predictions.
Here, $\operatorname{train}$ is the ``training'' data, which is $\{Y^{\mathit{OI}} \setminus
Y^V\}$ or $\{Y^{\mathit{OIA}} \setminus Y^V\}$,
depending on the model.
This can be viewed as an analogue of the frequentist procedure that
estimates the variance components and smoothing parameters by REML
(restricted maximum likelihood) and then solves the quadratic
minimization problem to fit the penalized spline.
Of course, rather than estimating the mode, the more time-consuming
alternative of estimating the posterior mean via MCMC could be used.
Treating $v$ and $\gamma$ as known (set to their estimated values)
allows us to derive the predictive distribution of the validation data
and prediction errors used for the prediction interval and probability
scores below [\citet{gn07}].

%
\begin{table}
\tabcolsep=0pt
\caption{Comparisons of cross-validation performance for the 8
candidate models
using averaged (over four subsets) criteria. Columns:
B---MSPE;
C---correlation;
D---empirical coverage of the prediction interval;
E---average width of the prediction interval;
F---negatively oriented interval score, equation (43) of \citet{gn07};
G---negatively oriented $\mathit{CRPS}$, equations (20)~and~(21) of \citet{gn07};
H---plug-in maximum likelihood prequential score, equation (54) of \citet{gn07}}\label{tab1}
\begin{tabular*}{\tablewidth}{@{\extracolsep{\fill}}@{}lccccccccc@{}}
\hline
\textbf{$\bolds{U}$: is}   & \textbf{$\bolds{\mathit{GST}}$: is}      & \textbf{$\bolds{A}$: is} \\ 
\textbf{$\bolds{u}$ used?} & \textbf{$\bolds{g_{\mathit{ST}}}$ used?} & \textbf{$\bolds{Y^A}$ used?} & \textbf{B} & \textbf{C} & \textbf{D} & \textbf{E} & \textbf{F} & \textbf{G} & \textbf{H}\\
\hline
0 & 0 & 0 & 0.264 & 0.674 & 0.878 & 1.359 & 3.336 & 0.955 & $-$0.083 \\
0 & 0 & 1 & 0.161 & 0.777 & 0.907 & 1.360 & 2.165 & 0.527 & $-$0.034 \\
0 & 1 & 0 & 0.580 & 0.610 & 0.840 & 1.338 & 6.184 & 2.376 & $-$0.175 \\
0 & 1 & 1 & 0.167 & 0.768 & 0.896 & 1.336 & 2.265 & 0.562 & $-$0.045 \\
1 & 0 & 0 & 0.143 & 0.802 & 0.937 & 1.402 & 1.975 & 0.435 & $-$0.004 \\
1 & 0 & 1 & 0.132 & 0.816 & 0.938 & 1.394 & 1.918 & 0.396 & \phantom{$-$}0.007 \\
1 & 1 & 0 & 0.172 & 0.759 & 0.906 & 1.383 & 2.267 & 0.543 & $-$0.035 \\
1 & 1 & 1 & 0.141 & 0.803 & 0.931 & 1.381 & 1.984 & 0.430 & $-$0.005 \\
\hline
\end{tabular*}
\end{table}

Once the predictions are available, we measure the predictive accuracy
using the mean squared prediction error (MSPE) and correlation between
the predicted and observed validation values (columns B and C in
Table~\ref{tab1}). We also considered criteria based on 95\%
prediction intervals, particularly the observed proportion of coverage
(column D), average width (column E) and the negatively oriented
interval score of \citet{gn07} defined as
\[
S_\alpha^{\mathrm{int}}(l,u;x) = (u-l) + \frac{2}{\alpha}(l-x)
\mathbb{I}(l-x) + \frac{2}{\alpha}(x-u)\mathbb{I}(x>u),
\]
where $\alpha= 0.05$, $l$ and $u$ are the lower and upper bounds of
the size $(1-\alpha)$ central prediction interval, and $\mathbb{I}(\cdot
)$ is the indicator function (column F). Given comparable empirical
coverages, lower values in columns E and F correspond to better fitting
models. Column G summarizes the negatively oriented continuously ranked
probability scores defined as
\[
\operatorname{CRPS}(F,x) = \int_{-\infty}^\infty\bigl\{F(y) -
\mathbb{I}(y \geq x)\bigr\}^2 \,dy,
\]
where $F(y)$ is the predictive distribution of interest, which has
recently drawn
the attention of the atmospheric sciences community [see \citet{gn07} and the references therein]. In column H we include
summaries based on the equivalent of the plug-in maximum likelihood
prequential score
$\sum_{j \in V_i} \log[Y^{(j)}|\operatorname{train}, v=\widehat{v},\gamma=\widehat
{\gamma}]$, where $Y^{(j)}$ is the $j$th observation in $V_i$. Higher
values in columns G and H correspond to better fitting models. The
criteria reported in the table are averaged over the four subsets of
data, for example,\vspace*{1pt} the average MSPE is $\sum_{i=1}^4 \|Y^{V_i} -
\widehat{Y}{}^{V_i}\|_2^2/n_i$, where $Y^{V_i}$ is the $i$th subset of
validation data, $\widehat{Y}{}^{V_i}$ is the corresponding vector of
predictions, and $n_i$ is the size of the validation subset.

The cross-validation results in Table~\ref{tab1} suggest that, for
every model
(and, actually, for every validation subset; see representative results
in Tables~4 and~5 in Section~A.3
of the online supplements [\citet{suppl14}])
inclusion of the \mbox{multiday} data through the linearized model---in order
to increase spatial coverage---always improves the predictive performance relative to the
corresponding model without multiday data. In particular, it can be
seen from Figures~8--11 in the online
supplements [\citet{suppl14}] that models with long-range interaction term $g_{\mathit{ST}}$
$(\mathit{GST}=1)$ do not perform well near the boundaries of the study region
if the model for $Y^A$ is excluded ($A=0$).

The two best models
are $M(U=1,\mathit{GST}=0,A=1)$ and $M(U=1$, $\mathit{GST}=1$, $A=1)$.
With an exception of one station where the model $M(U=1$, $\mathit{GST}=1$, $A=1)$
overpredicts, predictions from the two models are very similar,
suggesting that inclusion of the long-range spatio-temporal interaction
is not helpful for prediction given the observations available. It is
notable that, for the better models, the empirical prediction interval
coverage is close to the nominal 95\%. The small difference of 1--2\%
from the nominal coverage could be due to holding the values of the
parameters fixed at the estimated values. Failure to include the
short-range dependence term $u$ appears to result in underestimation of
the prediction error variance and, consequently, narrower intervals
with below-nominal coverage.

We also compared the predictive performance of our linearized models
and the corresponding ``simple models'' based on equation~(\ref
{eqref2}) proposed by a reviewer using the MSPE and the correlation
between held-out data and predictions. For each validation subset, our
linearized models outperformed the models of equation~(\ref{eqref2}).
Surprisingly, the na\"{\i}ve linearization of the ``simple models''
occasionally caused the predictive performance to deteriorate, relative
to the corresponding models without the multiday data. Our findings are
fully described in Section~A.3.2 in the online supplements [\citet{suppl14}].

Validating the model on a spatially and temporally disjoint subset of
data (online supplements [\citet{suppl14}], Section~A.3), which is indicative
of the models' out-of-sample prediction performance, yielded the same
choice of best model and the same conclusion that incorporation of the
aggregated data via linearization uniformly improves the quality of predictions.

\subsection{Assessment of the adequacy of linearization}\label{sublinvsnonlin}

Here we assess the impact of linearization on Bayesian inference using
models that include the multiday data based on the results of
Section~\ref{subpred}. We compare nonlinear and linearized versions of
model $M(U=0,\mathit{GST}=0,A=1)$ of Table~\ref{tab1} 
because the models with $(U=1)$ and/or $(\mathit{GST}=1)$ are computationally
less tractable.

Sampling from the linearized model was discussed in Section~\ref{seccasecomputationalmcmc}.
To sample from $[\gamma|Y^{\mathit{OIA}}]$ under the linearized model, we
initialized two Markov chains in a neighborhood of the mode of $[\gamma
|Y^{\mathit{OIA}}]$, sampling as discussed in Section~\ref{seccasecomputationalmcmc} for 75,000 iterations. The chains mixed
well, with lag-one correlations in the component-wise chains $\{\gamma
_j^{(i)}\}_i$ and $\{\log([\gamma^{(i)}|Y^{\mathit{OIA}}])\}_i$ around 0.95;
lag-one correlations between the corresponding components of
$v$ are of much smaller magnitude, typically between $0.2$ and $0.3$. A burn-in
sample of 2500 states was discarded from each chain.

To draw samples under the nonlinear model, we first reduced the dimension
of the posterior by analytically integrating out the vector $\alpha_0$.
We sampled from $[\gamma,w|Y^{\mathit{OIA}}]$ using the adaptive
RWMH sampler discussed in Section~\ref{seccasecomputationalmcmc}.
Here, we drew $\{w,\gamma\}$ in a single step when sampling from
$[w,\gamma|Y^{\mathit{OIA}}]$. This Markov chain mixes very slowly, with typical
lag-one correlations in $\{\log([\gamma^{(i)},w^{(i)}|Y^{\mathit{OIA}}])\}_i$ on
the order of 0.995. We used 6 Markov chains, each of length 200,000,
initialized in the high probability region of $[\gamma,w|Y^{\mathit{OIA}}]$. A burn-in
sample of 25,000 states was discarded from each chain. Based on the
effective sample size calculations, the Markov chain based on the
nonlinear model is about 10 times less efficient than the one based on
the linearized model.

%
\begin{figure}[b]

\includegraphics{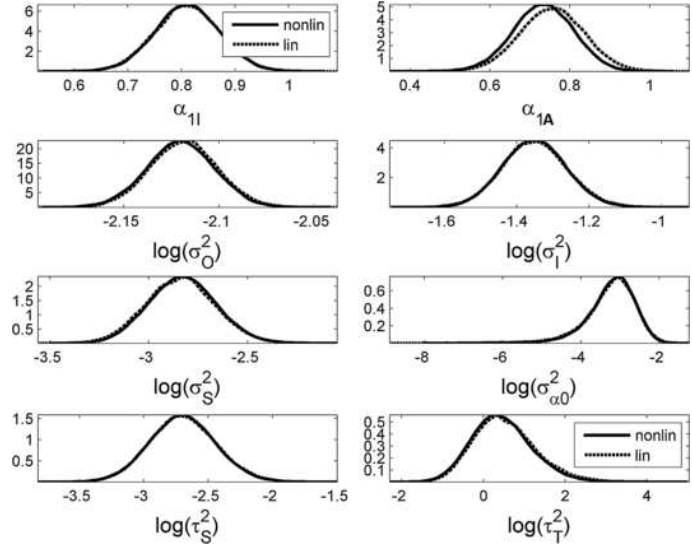}

\caption{Estimates of marginal densities of nonlinear parameters under
nonlinear (solid)
and linearized (dashed) versions of the model $M(U=0,\mathit{GST}=0,A=1)$.}\label{figlinnonlin}
\end{figure}

Estimates of the marginal posterior densities of $\gamma$ and $w$ are
shown in Figure~\ref{figlinnonlin}~and Figure~6
in the online supplements [\citet{suppl14}].
The marginal densities of elements of $\gamma$ and $w$ are remarkably
similar between the nonlinear and linearized models, with the exception
of the densities of $\gamma_2 = \alpha_{1A}$; these are still close to
one another. Plots of spatial predictions---obtained as means of the
posterior predictive distribution---for the two models (Figure~7 in the online supplements [\citet{suppl14}])
are also visually indistinguishable, which provides further support for
the use of linearization. The correlation between the spatial
predictions under the two models is 0.9999 on both the logarithmic and
original scale. We also examined the distribution and spatial
variability of the pointwise prediction differences. The variability of
the differences tends to increase with the distance from the central
site monitor (as expected due to the curse of dimensionality), but the
predictions are still very close to each other. The relative accuracy
of predictions on the original scale, computed pointwise as the
absolute value of the differences of the predictions divided by the
predicted value under the nonlinear model, is very high. For example,
the 90th, 95th, 99th and 99.5th percentiles for the empirical
distribution of the relative errors were 0.011, 0.014, 0.033 and 0.040,
respectively.

\subsection{Bayesian inference and prediction}\label{subbayes}

In this section we report results under the model chosen in Section~\ref{subpred}, which includes the short-range process, $u$, and the
multiday data but excludes the long-range $g_{\mathit{ST}}$ process.

To sample from $[\gamma|Y^{\mathit{OIA}}]$ using the computational strategy of
Section~\ref{seccasecomputationalmcmc}, we launched four Markov
chains, initialized in the region of high posterior probability of
$\gamma$. Each chain had a length of 12,500, and a burn-in sample of
size 2500 was discarded from each.
We examined trace plots of MCMC states and the corresponding posterior
density estimates to determine that the chains mixed rapidly and
converged to the same posterior.


%

%
\begin{table}[b]
\tabcolsep=0pt
\tablewidth=213pt
\caption{Posterior summaries of nonlinear parameters under model $M(U=1,\mathit{GST}=0,A=1)$}\label{tab3}
\begin{tabular*}{\tablewidth}{@{\extracolsep{\fill}}@{}lcccc@{}}
\hline
\textbf{Parameter} & \textbf{Mean} & \textbf{2.5\%} & \textbf{50\%} & \textbf{97.5\%}\\
\hline
$\alpha_{1I}$ & 0.956 & 0.870 & 0.957 & 1.040 \\
$\alpha_{1A}$ & 0.698 & 0.576 & 0.702 & 0.817 \\[2pt]
$\sigma_{O}^2$ & 0.045 & 0.041 & 0.045 & 0.049 \\[2pt]
$\sigma_{I}^2$ & 0.129 & 0.101 & 0.125 & 0.161 \\[2pt]
$\sigma_{S}^2$ & 0.037 & 0.021 & 0.035 & 0.056 \\[2pt]
$\sigma_{\alpha0}^2$ & 0.030 & 0.011 & 0.024 & 0.057 \\[2pt]
$\tau_S^2$ & 0.030 & 0.016 & 0.027 & 0.047 \\[2pt]
$\tau_T^2$ & 2.774 & 0.182 & 0.974 & 7.168 \\[2pt]
$\sigma_u^2$ & 0.098 & 0.090 & 0.098 & 0.106 \\
$\theta_S$ & 0.054 & 0.043 & 0.053 & 0.066 \\
$\theta_T$ & 0.120 & 0.073 & 0.120 & 0.169 \\
\hline
\end{tabular*}
\end{table}

Posterior means and quantiles for $\gamma$
are given in Table~\ref{tab3}. Even though parameters $\alpha_{1I}$
and $\alpha_{1A}$ have similar interpretations, $\alpha_{1A}$ is
smaller in magnitude than $\alpha_{1I}$. This suggests that multiday
indoor data are less informative for daily predictions of the outdoor
exposure process than daily indoor data. This is plausible because
readings from 30 out of 45 BCI sites overlap spatially and temporally
with those from BCO sites, whereas all BCA sites are spatially and
often temporally disjoint from the BCO sites. Consequently, the
spatio-temporal mismatch causes measurement error in the regressor (the
latent process here), and the coefficients are shrunk more toward zero
whenever there is more error in the covariate.
The temporal decay parameter $\theta_T \approx0.12$, interpreted in
light of the tapering structure, corresponds to a model with temporal
correlation function that is about 0.7 at lag one but decays faster
than that of the $\mathit{AR}(1)$ process with lag-one correlation of 0.7. The
spatial decay parameter $\theta_S \approx0.054$ (on the 1 km distance
scale) corresponds to spatial correlation that decays to 0.05 by about
35 km.
Consequently, when predicting within the temporal range of
measurements, the short-range process, $u$, ``pulls'' the predictions
toward the observed data, thereby capturing the nonperiodic features of
the exposure process not accounted for by $\zeta$.

\begin{figure}

\includegraphics{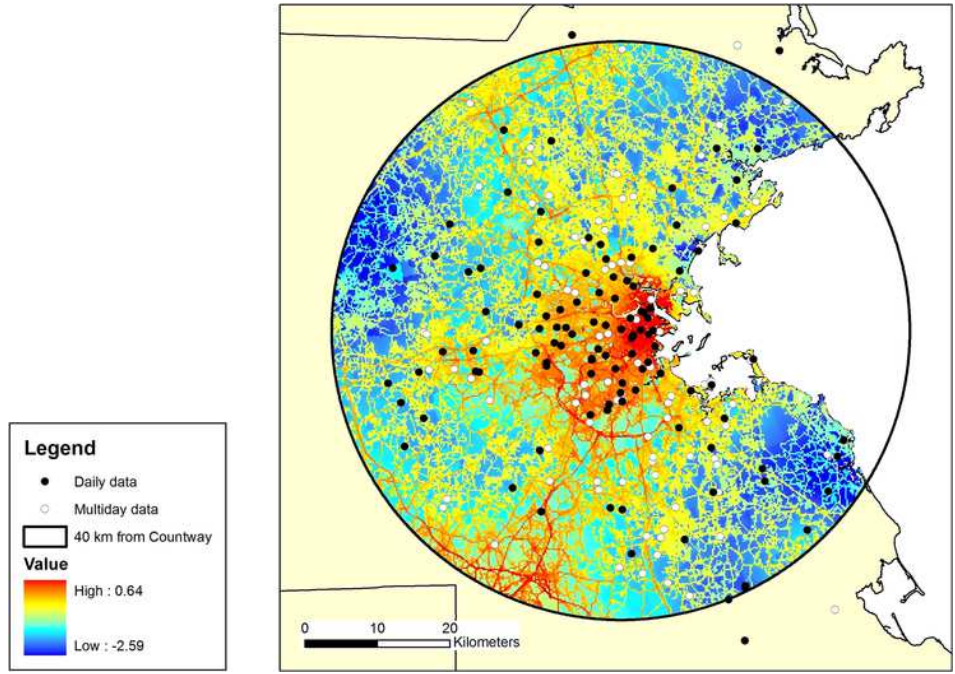}

\caption{Log black carbon predictions for July 31, 2006 based on the
mean of the predictive distribution, $E(\eta^P|Y^{\mathit{OIA}})$, under the
final model $M(U=1,\mathit{GST}=0,A=1)$. The unit is the natural logarithm of
$\mu g/m^3$.}\label{figfigcwetap}
\end{figure}

Posterior means and quantiles for the coefficients of the observable
covariates are reported in Table~\ref{tab4}. Based on preliminary
exploratory analysis using only the outdoor data, the logarithm of
readings from the central site monitor (\texttt{logHSPH}) was the most
important covariate for spatio-temporal prediction. From the Bayesian
model fit using data from all sources, this conjecture was further
supported by the relative widths and quantiles of the credible
intervals. The effect of other temporally-varying covariates such as
wind speed and the planetary boundary layer is not easily interpretable
in the presence of interactions of spatial and temporal covariates.
However, certain two- and three-way interactions have been shown to add
to the predictive ability of other prediction models in the Boston area
[\citet{zanob14}]. The spatially-varying population and land
use covariates are positively associated with the response. The traffic
density covariate is of most interest because of the relationship
between black carbon and traffic that motivates this work.
Its marginal effect---once the interactions with temporally varying
covariates have been accounted for---is positive, which can be clearly
seen in Figure~\ref{figfigcwetap}, in which predictions follow the
road network. In the early phase of this project we considered models
with fewer predictors and without interactions, which yielded a similar
relative ranking of the models and slightly less accurate predictions.

%

The primary goal of our work is to predict a vector of latent process
values, $\eta^P$, in the region for any temporal period of interest for
health effects analysis, which is done using $E(\eta^P|Y^{\mathit{OIA}})$.
Because of the temporal covariance tapering, if the minimum distance
between the temporal indices in $\eta^P$ and in $Y^{\mathit{OIA}}$ exceeds the
range of the taper function, then $E(\eta^P|Y^{\mathit{OIA}}) = E(C^P
w|Y^{\mathit{OIA}})$, where $C^P$ is the ``design matrix'' for $\eta^P$.
Figure~\ref{figfigcwetap} shows predictions for an example day (July~31, 2006) based on the MCMC estimate of $E(\eta^P|Y^{\mathit{OIA}})$.


\section{Discussion}\label{secdiscussion}

In this paper we developed a unified exposure prediction framework that
aggregates air pollutant concentration data from multiple disparate
sources that are available at different levels of temporal resolution,
which is of great importance for health effects models arising in
environmental science. We found that incorporation of even a modest
number of observations (93 or under~1.5\% of the overall observation
count) of the multiday data from a relatively large spatial network
(roughly doubling the number of unique spatial sites) uniformly
improves the prediction quality in a number of models that may or may
not include long-term and short-term spatio-temporal signal. To our
surprise, incorporation of a periodic long-range spatio-temporal trend
did not produce considerable improvements over the models without the
long-range interaction. We attribute this to the fact that the air
monitors in the networks corresponding to each study are scattered in
space and operate irregularly in time, which implies that the observed
data correspond to under 5\% of the dates from all the monitors over
the whole study period.
In our models, the departures from the periodic trend are being
captured by the short-range process. If the temporal coverage were
richer, we would be able to identify the nonperiodic component of
long-range variability better and to rigorously test its presence.

Our linearization approach provides a computationally efficient means
to build two quadratic approximations:
(i) the logarithm of $[\eta^A|v,\gamma,Y^{\mathit{OIA}}]$ and
(ii) the logarithm of $[\eta^A, \eta^P|v,\gamma,Y^{\mathit{OIA}}]$, where $\eta
^P$ is a vector
of latent process values we want to predict. This produces a linearized
model with Gaussian approximations for the marginal likelihood and
required conditional posterior densities.
An alternative that also results in Gaussian approximations is to
approximate (i) and (ii)
using a two-term Taylor expansion about the appropriate modes, which
needs to be located by a costly optimization run. Using these
approximations to integrate out
$\eta^A$ is equivalent to the Laplace approximation [\citet{tieka}]. The downside of
this scheme is that the approximation needs to be built for every value
of $(v,\gamma)$ of interest, which is infeasible in practice.

While we adopt an MCMC-based approach to Bayesian inference and prediction,
a promising direction for future work is to consider an approximation
scheme in the spirit of \citet{rueinla}, the integrated nested
Laplace approximation (INLA). Methodologically, one will need to
address the following two issues
that are critical to the computational performance of INLA for \emph{inference} in latent process models that combine multiple data sets,
both in our case study and in general.
First, one needs to be able to enforce the Markov property of the
spatio-temporal latent process. Second, increasing the number of data
sets in the joint model (that are linked by the latent process) and the
complexity of the model for the latent process adds to the dimension
of the hyperparameter vector~$\gamma$ [$\btheta$ in \citet{rueinla}],
which can make accurate numerical integration computationally
demanding, if feasible.

\section*{Acknowledgment}
We thank Steven Melly for help with figures.
The contents of this publication are solely the
responsibility of the grantee and do not necessarily
represent the official views of the U.S. EPA. Further,
the U.S. EPA does not endorse the purchase of any
commercial products or services mentioned in this
publication.

\begin{supplement}
\stitle{Supplement to ``Nonlinear predictive latent process models for\break integrating
spatio-temporal exposure data from~multiple sources''\break}
\slink[doi]{10.1214/14-AOAS737SUPP} 
\sdatatype{.pdf}
\sfilename{aoas737\_supp.pdf}
\sdescription{Online supplements contain technical details and supplementary figures and tables.}
\end{supplement}

%

\printaddresses
\end{document}